\documentclass[preprint2]{aastex}

\newcommand{\arcm}{\hbox{$^\prime$}}
\newcommand{\arcs}{\hbox{$^{\prime\prime}$}}
\def\degr{\hbox{$^\circ$}}
\def\fdg{\hbox{$.\!\!^\circ$}}
\newcommand{\mum}{\,\mu\mbox{m}}

\shorttitle{The F abundance in a Galactic Bulge giant}
\shortauthors{Uttenthaler et al.}

\begin{document}

\title{The fluorine abundance in a Galactic Bulge AGB star measured from
  CRIRES spectra\altaffilmark{1}}

\author{S. Uttenthaler\altaffilmark{\ddag}, B. Aringer, and T. Lebzelter}
\affil{Institut f\"ur Astronomie, University of Vienna,
  T\"urkenschanzstra\ss e 17, A -- 1180 Vienna, Austria}
\email{[uttenthaler;aringer;lebzelter]@astro.univie.ac.at}

\and

\author{H. U. K\"aufl and R. Siebenmorgen}
\affil{European Southern Observatory, Karl Schwarzschild Stra\ss e 2, D --
  85748 Garching near Munich, Germany}
\email{[hukaufl;rsiebenm]@eso.org}

\and

\author{A. Smette}
\affil{European Southern Observatory, Alonso de Cordova 3107, Vitacura,
  Casilla 19001, Santiago 19, Chile}
\email{asmette@eso.org}

\altaffiltext{1}{Data taken during CRIRES/ESO-VLT commissioning in June 2006.}

\altaffiltext{\ddag}{Current address: Instituut voor Sterrenkunde, K.\ U.\
  Leuven, Celestijnenlaan 200D, B-3001 Leuven, Belgium;
  stefan@ster.kuleuven.be}

\begin{abstract}
  We present measurements of the fluorine abundance in a Galactic Bulge
  Asymptotic Giant Branch (AGB) star. The measurements were performed using
  high resolution K-band spectra obtained with the CRIRES spectrograph, which
  has been recently installed at ESO's VLT, together with state-of-the-art
  model atmospheres and synthetic spectra. This represents the first fluorine
  abundance measurement in a Galactic Bulge star, and one of few measurements
  of this kind in a third dredge-up oxygen-rich AGB star. The F abundance is
  found to be close to the solar value scaled down to the metallicity of the
  star, and in agreement with Disk giants that are comparable to the Bulge
  giant studied here. The measurement is of astrophysical interest also
  because the star's mass can be estimated rather accurately ($1.4 \lesssim
  M/\mathrm{M}_{\sun} \lesssim 2.0$). AGB nucleosynthesis models predict only a
  very mild enrichment of F in such low mass AGB stars. Thus, we suggest
  that the fluorine abundance found in the studied star is representative for
  the star's natal cloud, and that fluorine must have been produced at a
  similar level in the  Bulge and in the Disk.
\end{abstract}


\section{Introduction}\label{intro}

Fluorine (F) is probably the element whose nucleosynthetic origin is
least known. The reason for the scarcity of knowledge is the fragility of its
only stable isotope ($^{19}$F) and the lack of measurable atomic lines in the
optical spectral range of normal stars. Apart from UV measurements of
highly ionized F in very hot stars \citep{Wer05} and optical measurements
of neutral F in extreme helium stars \citep{Pandey}, the only source of
information on stellar F abundances are vibration-rotation lines of the
hydrofluoric acid (HF). The HF molecule is efficiently formed in cool
stellar atmospheres (types $\sim$\,K0 and later), and a number of strong lines
appear in the near-IR $K$-band. The first identification of stellar HF lines
has been reported by \citet{Spi71}; lines of this molecule have been used by
\citet{Jor92} to measure F abundances outside the solar system for the first
time.

Several astrophysical sites for the synthesis of F have been proposed. It has
been confirmed by \citet{Jor92} through the correlation with the abundance of
carbon that AGB stars are indeed producers of F. This was also supported by a
recent study of F abundances in AGB stars of the LMC cluster NGC~1846
\citep{Leb08}. Whether or not AGB stars are
the {\it main} producers of F is still a matter of debate. \citet{MoMe}
estimate that AGB stars contribute to as much as 25\% of the Solar System
fluorine, but they emphasize the roughness of their estimate. \citet{Cun03} in
this respect rather suggest, based on observations of two red giants in
the globular cluster $\omega$~Cen, that $^{19}$F is created by neutrino
nucleosynthesis ($\nu$-process) during core collapse in supernovae of Type~II
\citep[SNe~II ][]{WoWe}. However, \citet{Federman} did not find a clear
indication for enhanced F abundance resulting from the $\nu$-process in a
region shaped by past supernovae. Finally, the Wolf-Rayet phase of massive
stars was investigated as a third site of F production by \citet{MeAr}.

The AGB phase is the final phase of nuclear processing for a broad initial
mass range between 0.8 and 8\,M$_{\sun}$. In this phase, the interior of the
star is structured in the following way: The inert carbon-oxygen core is
surrounded by a He-burning shell, on top of which is a He-rich shell, in turn
surrounded by a H-burning shell. The outer H-rich envelope is fully convective
and fills most of the star's volume. The He-burning shell ignites only
temporarily during so-called He-shell flashes or thermal pulses (TPs) to
release large amounts of energy. During a TP the He-rich shell gets thoroughly
mixed and is the site of rich nucleosynthesis (e.g.\ the s-process). For a
comprehensive review of AGB evolution see e.g.\ \citet{Hab04}. Among many
others, $^{19}$F is produced via complex reaction chains under these
conditions, or destroyed via $\alpha$ captures forming $^{22}$Ne. At very high
neutron densities, also n captures lead to a destruction of F. For a
discussion of the relevant reactions see \citet{For92}. The processed
matter is brought to the stellar surface via third dredge-up (TDU) mixing
events, which may operate after each TP.

Only few stellar systems have been investigated with respect to the abundance
of F: Besides {\it Galactic} field red giants \citep{Jor92}, similar stars
have been observed in the globular cluster M4 \citep{Smi05}, as well as in
$\omega$\,Cen and the LMC \citep{Cun03,Leb08}. \citet{Smi05} add F to the list
of elements known to vary in globular cluster stars and draw conclusions on the
early-cluster chemical pollution. \citet{Schuler} recently presented the
measurement of F in a very metal-poor star and found a considerable
overabundance of this rare element.

In this paper we report the measurement of the F abundance in a MS spectral
type Bulge AGB variable, i.e.\ an AGB star with an O-rich chemistry enriched
in s-process elements. It is the first measurement of this kind in a star of
the Galactic Bulge, and one of few F abundance determinations from a TDU-AGB
star. The observations have been carried out with the CRIRES infra-red
spectrograph.

\section{CRIRES Observations}\label{obs}

CRIRES is the CRyogenic Infra-Red Echelle Spectrograph mounted to the Nasmyth
focus A at the 8.2\,m Unit Telescope \#1 (Antu) of ESO's VLT on Cerro Paranal,
Chile. It has been installed at the telescope in 2006 followed by several
commissioning and science verification runs.
CRIRES is designed for high spectral resolution ($\lambda / \Delta \lambda $
up to $10^5$) and operates in the spectral range \mbox{0.95 $-$ 5.3\,$\mum$}.
A curvature sensing adaptive optics system feed is used to minimize slit losses
and to provide diffraction limited spatial resolution along the slit.
More details about the CRIRES instrument can be found in \citet{Kau04} and
on-line at \url{www.eso.org/instruments/crires/}.

One star observed during the first CRIRES commissioning run in June 2006 was
the Bulge Mira variable \object[Plaut 3-1347]{M1347} \citep{Wes87}, also
known as Plaut~3-1347 or V2017~Sgr (J2000 coordinates: 18$^{\rm h}$ 38$^{\rm
  m}$ 45\hbox{$.\!\!^{\rm s}$}7  $-$34\degr\ 33\arcm\ 28\arcs). With a
Galactic latitude of about $-$12\fdg6 it belongs to the outer Bulge. The
observations of M1347 were carried out on June 07, 2006. A large part of the
near-IR K-band was observed, although the analysis presented here is limited
to two wavelength settings covering the range \mbox{2.253 $-$ 2.310\,$\mum$}.
The slit-width was set to 0\farcs2; thus, the maximum resolution of 100\,000
(3\,km\,s$^{-1}$ equivalent) was achieved. The integration time was 60\,s for
each of the four nodding positions per setting. A hot standard star at similar
airmass was observed immediately afterwards. The raw frames were reduced with
the CRIRES pipeline (version 0.2.3), and the 1D science and standard star
spectra were wavelength-calibrated separately using the numerous telluric
absorption lines present on all of the four detector arrays. The
wavelength-calibration was done separately for the science and telluric
standard star spectrum because of the limited reproducibility of the Echelle
grating position. Finally, the science spectrum was divided by the standard
star spectrum to correct for the telluric lines and the illumination pattern
as well as possible. Note that the telluric lines are strong enough to use
them as wavelength calibrator, but they are weak enough to be corrected for by
standard star division; thus they have no influence on the abundance
measurements presented here.
The signal-to-noise ratio was estimated from an overlapping region on chip
\#3 that was observed in both settings to be close to 100.

\section{Stellar parameters}\label{params}

M1347 has previously been observed together with a larger sample of Bulge AGB
stars with the UV and Visual Echelle Spectrograph (UVES) at the VLT on July
08, 2000 (= JD 2451733.6) in the blue arm (\mbox{377 $-$ 490\,nm}) and in the
red arm (\mbox{667 -- 1000\,nm}). It was found to show absorption lines of
technetium \citep{Utt07a} and lithium \citep{Utt07b}. Tc is an indicator of
recent or ongoing s-process and TDU in an AGB star. As probably all genuine
Bulge AGB variables this star is O-rich. Strong absorption bands of metal
oxides such as TiO, VO, and ZrO are evidence that M1347's C/O ratio is below
unity. Because of the O-rich chemistry and signs of s-process enrichment
(Tc, enhanced ZrO bands) it is classified as type MS \citep{Ste84}. The
(moderate) Li content of
\mbox{$\log \epsilon(\rm{Li}) = +0.8$}~\footnote{The abundance of element X on
this scale is given by
$\log \epsilon(\mathrm{X}) = \log N(\mathrm{X})/N(\mathrm{H}) + 12$.} in this
star probably results from a different mixing phenomenon called
{\it cool bottom processing} \citep[CBP, e.g.\ ][]{Was95}, although other
enrichment scenarios can not be completely excluded \citep{Utt07b}. 

The mass and luminosity of M1347 are rather well constrained, thus its
abundances can be easily compared with expectations from models of stellar
evolution and nucleosynthesis. From Fig.\ 3 of \citet{Str03} the mass can be
limited to the range $1.4\,\rm{M}_{\sun} < M < 2.0\,\rm{M}_{\sun}$ over a wide
range in metallicity ($0.15\,\mathrm{Z}_{\sun} < Z < \mathrm{Z}_{\sun}$).
In the Bulge, due to its age, no high-mass stars
\mbox{($\gtrsim 2\,\mathrm{M}_{\sun}$)} reaching \mbox{C/O$>$1} because of
dredge-up of carbon on the TP-AGB are present anymore. Below roughly
1.4\,M$_{\sun}$ no dredge-up occurs at all, preventing a star to become
s-process enriched.
Note that the former constraint would not hold for a Disk star, since it then
could be a higher mass star on its way to \mbox{C/O$>$1} experiencing one of
its first TDU events on the TP-AGB.
Additional constraints come from the \mbox{$M_{\rm{bol}} - P$} diagram. Using
linear pulsation models \citep{WoSe96} M1347 is placed slightly above
1.5\,M$_{\sun}$ \citep[see Fig.\ 5 in ][]{Utt07a}. The luminosity at a distance
of 8.0\,kpc is measured to be \mbox{$L \cong 11\,700\,\rm{L}_{\sun}$}
(\mbox{$M_{\rm{bol}} = -5\fm43$}), and the period is 426.26\,d
\citep{Utt07a,Utt07b}.

The CRIRES observations on JD 2453895.9 were carried out 5.07 light cycles
after the UVES observations. We thus assume that, despite the star's
variability, M1347's atmosphere was in a very similar state at the time of
CRIRES observations compared to the UVES observations. We estimate atmospheric
parameters from the UVES spectra by comparing them to a grid of synthetic
spectra based on COMARCS atmospheric models. COMARCS is a revised version of
MARCS \citep{Joe92} with spherical radiative transfer routines from
\citet{Nordlund} and new opacity data from the COMA program \citep{Ari00}.
Similar to the analysis of \citet{Gar07} a $\chi^2$ minimization method
was used to find the parameters of the model best fitting the TiO
$\gamma$(0,0)Ra (705.6\,nm) and $\gamma$(0,0)Rb (709.0\,nm) band heads in the
UVES spectra \citep{Utt07b}. The parameters found for M1347 are
$T_{\rm{eff}} = 3200$\,K, $\log g = -0.5$, [M/H]~=~0.0. Furthermore, the
parameters $\xi = 3.0\,\rm{km\,s}^{-1}$, $M = 1\,\rm{M}_{\sun}$, C/O~=~0.48,
and [Ti/H]~=~$+0.2$ have been adopted for the calculation of the model grid.
This model atmosphere was used to refine the metallicity of the star M1347 as
explained in the next section. (Note here that the stellar mass assumed for the
model atmosphere has a negligible effect on the spectrum and therefore is no
indication of the real mass of the star.) The error on the effective
temperature, the parameter on which abundance analysis is most sensitive, was
estimated from the $\chi^2$ analysis to be $\pm 100$\,K. The spectral type of
M1347 was determined from the UVES spectrum using the relations of
\citet{Fluks} as M7S. On the effective temperature scale of \citet{Ridgway},
this corresponds to $T_{\rm{eff}} = 3126$\,K, well inside the estimated error
range. Assuming 1.5\,M$_{\sun}$ for the mass and $\log g = -0.5$ and applying
the Stefan-Boltzmann law to the measured luminosity results in a surface
temperature of 3158\,K.

\begin{figure*}
  \centering
  \includegraphics[width=15.0cm, bb = 95 370 541 703, clip]{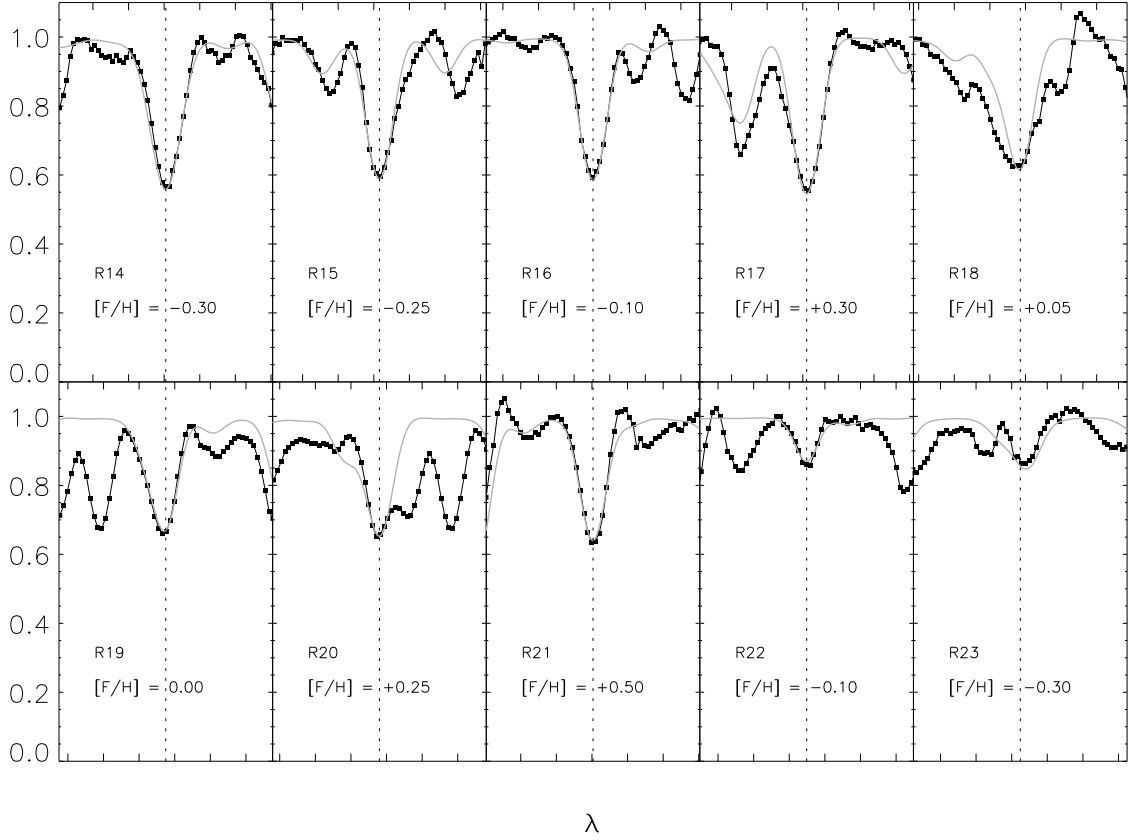}
  \caption{The R14 to R23 lines of HF in the Bulge giant star M1347 (black line
    with dots) and their best-fit model spectrum (grey line), with the
    corresponding F abundance given in the legend. The central wavelength,
    marked as vertical dashed line, is given in Table~\ref{HF_lines}. The tick
    marks on the x-axis are 0.1\,nm apart.}
  \label{M1347HF}
\end{figure*}

\section{Analysis}\label{analysis}

\subsection{Spectral synthesis}\label{subspectsynth}

The lines R12 to R28 of the vibrational 1$-$0 band of the HF molecule fall in
the observed spectral range. The lines R12 and R13 are, despite the high
resolution, hopelessly lost in the $^{12}$CO 2$-$0 band (head at
$\sim$\,2293.5\,nm) and cannot be used for abundance analysis. Lines above
R23 are very weak and could not be identified. The lines blue-ward of the
$^{12}$CO 2$-$0 band head however are well suited for the analysis. We want to
note here that all previous abundance analyses based on HF, except that of
\citet{Jor92}, rely on a single line of that molecule!

A calculated line list of the HF molecule was kindly provided by R.~H.\ Tipping
(priv.\ comm.). This list is unpublished, and has been used previously by
\citet{Jor92} for the measurement of F abundances in Disk giants. The $\log gf$
values of the lines could be derived from the Einstein $A$ values provided in
the Tipping list with the corresponding standard transformation
\citep[as described in ][]{Bernath} assuming a rotational degeneracy of
2J$'' + 1$. Together with the as well provided excitation potential, the data
can be used for spectral synthesis calculations. Additionally, a measured line
list by \citet{Web68} is available which is also incorporated into the HITRAN
data base \citep{Rot05}. For the 1$-$0 band of HF, this list contains only
lines up to R14. We performed a comparison based on synthetic spectra between
these lists for the lines of the 1$-$0 band in common and found very good
agreement, both in wavelength and line strength. We thus assume that the error
in the abundance analysis (section~\ref{errana}) resulting from uncertainties
in line data is negligible. The calculated list of Tipping can be regarded as
quite accurate. Since the Tipping data reaches to higher rotation quantum
numbers, we only used the data from that list for the abundance analysis.

\begin{table*}
\caption{Lines of the HF molecule identified in the star M1347. The values of
  the wavelength, $gf$, and $\xi$ are derived from the calculated list of
  R.~H.\ Tipping (priv.\ comm.).}
\label{HF_lines}
\begin{tabular}{cccccc} \\
\hline \\
Line & $\lambda$ (nm) & $gf (10^{-5})$ & $\xi (10^3 \mathrm{cm}^{-1}$) & EW
(mnm) & best-fit [F/H] \\
\hline \\
R14 & 2289.305 & 11.41 & 6.276 & 43.5 & $-0.30$\tablenotemark{a}  \\
R15 & 2283.316 & 11.18 & 6.865 & 41.2 & $-0.25$\tablenotemark{a}  \\
R16 & 2278.453 & 10.87 & 7.489 & 35.9 & $-0.10~$            \\
R17 & 2274.711 & 10.47 & 8.148 & 47.3 & $+0.30$\tablenotemark{a}  \\
R18 & 2272.085 & 9.993 & 8.841 & 55.7 & $+0.05~$            \\
R19 & 2270.575 & 9.449 & 9.566 & 30.1 & $~~0.00$\tablenotemark{a} \\
R20 & 2270.180 & 8.843 & 10.32 & 26.3 & $+0.25$\tablenotemark{a}  \\
R21 & 2270.903 & 8.185 & 11.11 & 30.5 & $+0.50~$            \\
R22 & 2272.749 & 7.483 & 11.93 & ~9.8 & $-0.10~$            \\
R23 & 2275.724 & 6.748 & 12.78 & 11.5 & $-0.30$\tablenotemark{a}  \\
\hline
\end{tabular}
\tablenotetext{a}{\protect Line is observed in both spectrograph wavelength
  settings, and the mean of the EW and abundance determination is given.}
\end{table*}

The TiO band heads used for the temperature determination are also somewhat
sensitive to the metallicity [M/H] of the star, leading to a temperature $-$
metallicity degeneracy. We thus attempted to check the metallicity that was
found by the $\chi^2$ minimization using atomic lines situated in the range
observed with CRIRES. The reliability of $\log gf$ values of atomic lines in
the relevant wavelength range contained in the VALD database \citep{Kup99} was
checked by comparing an observed \object{Arcturus} spectrum \citep{Hin95} with
a corresponding synthetic COMARCS spectrum. Four lines turned out to be well
suited for the metallicity determination: a Na line at 2208.969\,nm, a Ca line
at 2282.718\,nm, and two Fe lines at 2274.514 and 2283.870\,nm,
respectively. From these lines we determine a metallicity of [M/H] =
$-0.17^{+0.14}_{-0.22}$, where the error comes from the scatter among the
values from the individual lines (partly covered by two spectrograph
settings). In principle, one could iterate the procedure to find the
temperature of the star assuming a metallicity of \mbox{[M/H] = $-0.17$},
and with that temperature iterate again on the metallicity, etc. However, we
find from our models that a change of $-0.5$\,dex in [M/H] has the same effect
on the TiO band strength as a change of about $-100$\,K in $T_{\rm{eff}}$,
which is the accuracy of our temperature estimate. Thus, the impact on the F
abundance measurement should be well within the estimated error bars
(section~\ref{errana}). Nevertheless, the found metallicity agrees well with
the peak metallicity of Bulge stars \citep[\mbox{[M/H] = $-0.1$}; ][]{Zoc03},
and is just slightly below solar metallicity. We calculated a COMARCS
atmosphere with the metallicity of [M/H] = $-0.17$ and used it for the
following spectral synthesis calculations.

The best-fit F abundance was determined individually for the ten identified
and measurable HF lines in the observed spectral range. Spectral synthesis was
done in steps of 0.05\,dex in F abundance. A solar reference value
$\log \epsilon(\mathrm{F}) = 4.56$ was adopted \citep{Gre98}.
Table~\ref{HF_lines} lists the identified lines with their calculated
wavelength, $gf$ value, excitation potential $\xi$, measured equivalent width
(EW), and the best-fit [F/H] value derived from the respective line. In the
spectral synthesis calculations, all atomic species and molecular lines from
CO, CH, C$_2$, SiO, CN, TiO, H$_2$O, OH, VO, CO$_2$, SO$_2$, HCl, CH$_4$, FeH,
CrH, and ZrO were included to account for blending lines as well as possible.
The line data references are summarized in Table~1 of \citet{Cristallo}. An
additional macroturbulence of $v_{\rm{macro}} = 3.0\,\rm{km\,s}^{-1}$ has been
added in the spectral synthesis. For best-fit model spectra see
Fig.~\ref{M1347HF}.

From Table~\ref{HF_lines} it becomes obvious that the F abundances required
to fit the different HF lines have a large spread. This is most probably
because some of the observed lines are blended, and the respective blending
line is not adequately represented in any of the used line lists. There exists
a general problem of uncertain or simply incomplete (molecular) line data in
terms of wavelength and/or $gf$ value which becomes particularly apparent at
the high resolution of the present spectra. This difficulty is known, but
detailed comparison to observed data is scarce \citep[][ is one of few
examples]{Ari05}. In the wavelength region analysed here, line data are
uncertain especially for the CN list (see below). Also, dynamic and mass loss
effects may lead to the formation of spectral lines which are not formed in
hydrostatic atmosphere models used here \citep{Now05}.

The rotational lines inside a certain vibrational band of di-atomic
molecules reach a maximum strength at a certain rotation quantum number,
depending on the gas temperature. For the HF 1$-$0 band, this maximum is
reached at the R5 line for the temperatures of the HF line forming layers. For
higher and lower rotation quantum numbers the line strengths (opacities) will
decrease monotonically. Some lines in our observed spectrum clearly deviate
from this trend; these are the R17, R18, R20, and R21 lines. Although at least
the R17 and R21 lines do not appear asymmetric, which would definitely hint at
their blended nature, a blend of unknown origin and strength is assumed for
these lines. The R18 and R20 lines have a rather asymmetric profile and are
thus certainly blended. The R19 line is blended with a CN line in the
synthetic as well as in the observed spectrum, but the contribution by CN
seems to be well modelled. CN has been discarded from the list of molecules
used for the spectral synthesis of the R22 line because a CN feature appeared
in the red wing which is not present in the observed spectrum. Also R23, the
highest-lying detected line, is problematic. It is blended with a CN line,
too, with only a minor contribution by HF. We thus decided to exclude the
lines R17, R18, R20, R21, and R23 from the determination of the average F
abundance. The average F abundance from the remaining five lines in this
``cleaned'' list (R14, R15, R16, R19, and R22), calculated by converting the
individual F abundances to a linear scale before averaging them and converting
the average back to the logarithmic scale, is \mbox{[F/H]= $-0.14$}. The
differences between observed and synthetic EWs of these lines, weighted with
the inverse of their EW, reach least squares at \mbox{[F/H]~=~$-0.16$}.

We note that there is perhaps a trend of increasing [F/H] with increasing
rotation quantum number even in the cleaned list of lines. Due to the very
good agreement in terms of wavelength and gf values between the HITRAN and the
Tipping line data for the lines up to R14 we regard incorrect line data as the
source of this trend as unlikely. We performed abundance measurements with a
model atmosphere with an effective temperature increased by 200\,K (i.e.\
3400\,K); however, the trend remained. Thus, the trend is obviously not
connected to a wrong temperature estimate. The only solution we can offer at
the moment is that the trend might be related to a deviation of the
temperature structure of the model atmosphere from that of the real
atmopshere, but this has to be regarded as tentative. Whatever the reason for
the trend is, we again want to stress that our study is only the second which
employs more than one line to derive F abundances from HF lines.

\subsection{Error analysis}\label{errana}

Determining abundances in a Bulge Mira star is a tricky task. Unlike for
globular cluster stars \citep[e.g.\ ][]{Smi05}, no a priori value of the
metallicity is known, and the temperatures of cool, pulsating Mira variables
are difficult to measure precisely, not to mention other stellar parameters.
The uncertainty on the F abundance derived here, as shown below, is higher than
found in other studies \citep[e.g.\ ][]{Smi05}.

A large source of error, due to its effects on molecule dissociation and
level population, is the uncertainty in the effective temperature. By varying
the temperature of the model atmosphere by $\pm 100$\,K, which is a good
measure of the temperature uncertainty, we find a variation of [F/H] of
$\pm0.15$\,dex from all lines in the cleaned list.

The metallicity [M/H] has an impact on the measured F abundance comparable
to that of the effective temperature because of its influence on the
atmospheric structure. We therefore calculated spectra at the metallicities of
the upper and lower limits of the metallicity range of M1347 (steps of
$^{+0.14}_{-0.22}$ to the adopted value of \mbox{[M/H] = $-0.17$}). In fact, we
found that the F abundance changes in lockstep with changes in the metallicity,
which translates into an uncertainty of the F abundance of the same magnitude
as the uncertainty in [M/H]. The HF lines become stronger upon decreasing the
overall metallicity while keeping the F abundance constant, thus a lower
metallicity of the model atmosphere results in an F abundance lowered by the
same amount.

The surface gravity $\log g$ has only a minor effect on the derived F
abundance. Increasing $\log g$ from $-0.5$ to $0.0$, a range generally occupied
by red giant stars, leads to an F abundance reduced by 0.04\,dex.

The values of the macro- and micro-turbulent velocities, $v_{\rm{macro}}$
and $\xi$, could be checked for from the profiles of the HF lines in the
cleaned list. A value of $3\,\mathrm{km\,s}^{-1}$ for both of them was found
to very satisfyingly reproduce the profiles. A deviation of the sum of
$v_{\mathrm{macro}} + \xi$ by more than $2\,\mathrm{km\,s}^{-1}$ leads to line
profiles noticeably different from the observed profiles. This maximum
deviation leads to an uncertainty of [F/H] of $\pm 0.1$\,dex.

The specific value of the C/O ratio, which may differ from the solar value
due to first and third dredge-up processes, may have an influence on the F
abundance measurement in three different ways. First, it influences the
strength of the TiO bands used for the temperature determination. Second, the
atmospheric structure depends on the C/O ratio. Third, the C/O ratio influences
the strength of lines blended with the HF lines, in particular of the very
C/O-sensitive CN lines.

From a number of CN lines falling in the wavelength range observed with CRIRES
which are identified in the Arcturus spectral atlas of \citet{Hin95}, we
estimate that the C/O ratio in the atmosphere of M1347 is around 0.7. Since
with the present data we cannot have an independent check on the O and N
abundance, solar O and N abundances were assumed in this estimate. The
occurrence of lithium in M1347 has been ascribed to CBP acting in this star
\citep{Utt07b}. If this interpretation is correct, the N abundance could be
possibly enhanced because some $^{12}$C will be converted to $^{14}$N by CN
cycling. We expect that the effect of this shift on the CN lines is a minor
one.

Because the CN lines in the observed spectrum can by far be not as well fitted
as the HF lines, our estimate of the C/O ratio is quite rough. For this reason,
we decided to not change the C/O ratio of the model atmosphere that was used
for the abundance analysis from its adopted value of 0.48. We rather estimate
from the difference to \mbox{C/O = 0.7} what the effect on the measurement of
the F abundance is. We found from our model atmospheres that an increased C/O
ratio {\it increases} the strength of TiO bands (which were used for the
temperature determination). A naive expectation would be a decrease in TiO band
strength because less O is available to form TiO as C/O is increased. The prime
effect of an increased C/O ratio however is the same as an increased
metallicity, namely that the atmospheric structure is changed such that the TiO
bands become stronger. Only at a C/O ratio of 0.9 or higher, the effect of
a decreasing number of oxygen atoms available to form TiO takes over, and the
TiO bands finally decrease in strength. We find from our models that a C/O
ratio of 0.7 implies a temperature of M1347 increased by about 30\,K (compared
to \mbox{C/O = 0.48} , which in turn implies a F abundance {\it increased} by
$\sim 0.05$\,dex (see above). The effect on the atmosphere structure implies an
F abundance {\it decreased} by $\sim 0.05$\,dex.  In our cleaned list of HF
lines, only the R19 line is blended with a weak CN line, but its effect on the
F abundance measured from the R19 line is negligible. Thus, the effects on the
F abundance measurement of an increased C/O ratio (via the temperature
determination on the one hand and the changed atmosphere structure on the other
hand) effectively cancel out.


The error due to the (insecure) continuum placement can be estimated
from lines observed in both wavelength settings; it amounts to $\pm0.05$\,dex.
Finally, we include the statistical standard deviation of the F abundance
derived from the individual lines of the cleaned list, which is
$^{+0.11}_{-0.14}$\,dex, again calculated on a linear scale.

Converting the individual error items to a linear scale before summing in
quadrature gives a total error of $^{+0.23}_{-0.28}$\,dex. The use of a
hydrostatic model atmosphere for a variable star may induce an additional
systematic error which we cannot estimate here. Table~\ref{errors} summarizes
the sources of error and their magnitudes.

\begin{table}[!th]
\caption{Individual sources of error and their magnitude. The statistical
spread is calculated only on the basis of the lines in the cleaned list.}
\label{errors}
\begin{tabular}{cc} \\
\hline \\
Source      & $\Delta$ [F/H] \\
(Parameter) & (dex)          \\
\hline \\
$T_{\mathrm{eff}}$         & $\pm0.15$ \\
$\log g$                   & $\pm0.04$ \\
$\xi + v_{\mathrm{macro}}$ & $\pm0.10$ \\
$[\rm{M}/\rm{H}]$          & $+0.14 / -0.22$ \\
C/O                        & $\pm0.00$ \\
continuum                  & $\pm0.05$ \\
stat. standard dev.        & $+0.11 / -0.14$ \\
Total                      & $+0.23 / -0.28$ \\
\hline \\
\end{tabular}
\end{table}

\section{Conclusions and Outlook}\label{conclusio}

We measure a fluorine abundance in the Galactic Bulge AGB star M1347 of
[F/H] = $-0.14^{+0.23}_{-0.28}$. With respect to the metal abundance we find
[F/M] = $+0.03^{+0.20}_{-0.21}$ (the error bar is smaller because [F/H] changes
in step with changes in [M/H], thus the uncertainty in [M/H] does not enter the
errorbudget of [F/M]). The F abundance in M1347 is thus in agreement with the
scaled solar abundance. Note, however, that the solar system meteoritic F
abundance is somewhat insecure, and reduced with respect to the F abundance in
normal K-M giants of the solar neighbourhood \citep[cf.\ ][]{Jor92}.

The F abundance measured in M1347 can be compared to Disk stars analysed
by \citet{Jor92}. Of their list, the stars Y~Lyn and RS~Cnc are particularly
comparable to M1347: Both have oxygen-rich atmospheres with positive
detections of Tc, are of late spectral type (for Y~Lyn the classifications
M5Ib-II and M6S are found in the literature, for RS~Cnc the classification is
M6IIIS), and the effective temperature is quoted with 3200\,K by
\citet{Jor92}. Note, however, that Y~Lyn and RS~Cnc are semi-regular variables
with a pulsation period of only 110 and 120\,d, respectively, compared to the
Mira-like variability of M1347 with a period of 426.6\,d. This may hint
towards a somewhat higher mass of Y~Lyn and RS~Cnc compared to that of M1347
\citep{LH99}. \citet{Jor92} determined F abundances of [F/H]~=~+0.15 for
Y~Lyn, and [F/H]~=~+0.13 for RS~Cnc, respectively, and quote error-bars of the
order of 0.2 -- 0.3\,dex. The F abundance of these two disk stars and the
Bulge star analysed here thus overlap within the error bars.

The mass of the Bulge giant M1347 is estimated to be in the range
$1.4 \lesssim M/\mathrm{M}_{\sun} \lesssim 2.0$, probably close to
1.5\,M$_{\sun}$. The measured F abundance can thus be compared to theoretical
predictions of AGB evolutionary and nucleosynthesis calculations. Predictions
of the F surface abundance for a wide range of stellar masses and
metallicities is presented e.g.\ by the calculations of \citet{Kar03}. Their
lowest mass models experiencing TDU have $M = 2.25\,\mathrm{M}_{\sun}$ for $Z =
0.02$ and $M = 1.75\,\mathrm{M}_{\sun}$ for $Z = 0.008$. Figures~C7 and C18 of
\citet{Kar03} include the evolution of the surface abundance of $^{19}$F of
these models. For the mentioned combinations of mass and metallicity, the
surface abundance of F is enhanced only by $\sim 0.1$\,dex on the TP-AGB due to
internal nucleosynthesis and third dredge-up. A considerable increase is
predicted only at higher masses
\citep[$\sim 3\,\mathrm{M}_{\sun}$, see also ][]{Lug04}. Trusting in these
model predictions, we may assume that the F we see in M1347 increased at most
by $0.1$\,dex from its initial abundance, i.e.\ the F abundance of M1347's
natal cloud.

It is interesting to compare this inferred Bulge fluorine abundance to
predictions of Galactic chemical evolution models. A semi-analytic multizone
chemical evolution model of fluorine in the Milky Way has been presented by
\citet{Renda}. The authors find satisfactory agreement between their model and
the F abundance in solar neighborhood stars only if the contributions by
Wolf-Rayet and AGB stars are taken into account. The F abundance in
$\omega$~Cen giants however is well fit by F production in SNe~II alone. In a
similar manner, our measurement of the F abundance in the Bulge giant M1347,
which is in agreement with the solar abundance, suggests that Wolf-Rayet and
AGB stars also played a significant role in the chemical evolution of the
Galactic Bulge. The formation of the Bulge then must have lasted long enough
that its chemical evolution could have been influenced by these stellar types.
We thus conclude that Wolf-Rayet and AGB stars contributed to the chemical
evolution of the Galactic Bulge and Disk on a very similar level.

\citet{Lug04} speculate that extra mixing processes (CBP) that connect the
convective envelope to the H-burning shell might be the key to understand the
high F abundances at a given C/O ratio as observed by \citet{Jor92}. The
occurrence of Li in the Bulge star analysed in the present study has been
ascribed to the operation of CBP, too \citep{Utt07b}. However, the effect on
CBP is rather that of decreasing the C/O ratio (because of the transformation
of C into N, see also section~\ref{errana}) than influencing the F abundance.
If layers with temperatures exceeding some 30 million~K, also some F would be
destroyed via $^{19}$F(p,$\alpha$)$^{16}$O. It is not clear from current
theoretical and observational studies if such high temperatures are involved in
CBP an if destruction of F could indeed happen. Also from our observations no
clear conclusion in this respect can be drawn. For further details on the
connection between CBP and the fluorine abundance we refer to \citet{Lug04}.

It would be of interest to compare the F abundance in Bulge objects without
signatures of internal nucleosynthesis (e.g.\ stars on the first giant
branch) to the F abundance derived in the present study. If the F abundance in
such stars is found to be close to our result, a confirmation of the
predictions of detailed nucleosynthesis calculations such as those of
\citet{Kar03} for low mass stars is suggested. With the advent of
high-resolution IR spectrographs like CRIRES such measurements have become
readily feasible.

A deeper investigation of the nucleosynthetic origin of F is desired. We are
currently devising a CRIRES observing program aimed at determining the F
abundance in high-mass AGB stars of the Magellanic Clouds. In this respect
the somewhat elaborate error analysis in section~\ref{errana} is a useful
and important exercise. The goal of these CRIRES observation will be to observe
the predicted up-turn of F production at masses \mbox{$\sim3\,\rm{M}_{\sun}$},
and the F {\it destruction} at masses around 5\,M$_{\sun}$ and
higher. Important constraints on high-mass AGB evolution and reaction rates
involved in the synthesis of $^{19}$F can be expected from these measurements.

\acknowledgments

We particularly acknowledge R.~H.\ Tipping for providing his unpublished line
list of the HF molecule. We also thank A.\ Jorissen for valuable comments on
the manuscript. SU and TL acknowledge funding by the Austrian Science
Fund FWF under the project P 18171-N02, and BA acknowledges funding by the FWF
under project P 19503-N13. We want to thank the CRIRES commissioning team for
carrying out the observations of the Bulge star analysed here.

{\it Facilities:} \facility{ESO VLT (CRIRES)}

\end{document}